# Triple-Poles Complementary Split Ring Resonator for Sensing Diabetics Glucose Levels at cm-Band

Ala Eldin Omer[1,2], George Shaker[1], Safieddin Safavi-Naeini[1], Georges Alquié[2], Frederique Deshours[2], and Hamid Kokabi[2]
[1]Centre for Intelligent Antenna and Radio Systems (CIARS), University of Waterloo, Waterloo, Canada
[2]Laboratory of Electronics and Electromagnetism (L2E), Sorbonne University (SU), Paris, France

*Abstract*—Microwave sensors are very promising for sensing the blood glucose levels non-invasively for their non-ionizing nature, miniaturized sizing, and low health risks for diabetics. All these features offer the possibility for realizing a portable non-invasive glucose sensor for monitoring glucose levels in real time. In this article, we propose a triple poles complementary split ring resonator (CSRR) produced on a FR4 substrate in microstrip technology in the cm-wave band (1 – 6 GHz). The proposed bio-sensor can detect the small variations in the dielectric properties (relative permittivity and dielectric losses) of glucose in the blood mimicking aqueous solutions due their intense interaction with the electromagnetic field at harmonic resonances. The resonator exhibits higher sensitivity performance at the different resonances compared to the single and double-poles counterparts as demonstrated by simulations in a 3D full-wave EM solver.

*Keywords—microwave bio-sensor; split-ring resonator; non-invasive glucose sensing; Debye model*

## I. Introduction

Diabetes is directly caused by the breakdown of steady insulin production, thus affecting a cell's ability to absorb glucose from the bloodstream. Over 300 million people worldwide are involved with Diabetes and the number is increasing rapidly at an unpredictable rate every year according to the International Diabetes Federation (IDF) report [1]. To meet glycemic targets and to avoid the susceptibility of any serious complications like heart disease, stroke, coma, kidney failure, blindness, etc … [2], diabetics are recommended to check their blood glucose levels 2-4 times per day and control it as necessary through medication [3][4]. Generally, typical glucose levels in human beings blood for type 2 diabetes widespread among 90% of people normally vary in the range from 72 to 140 mg/dL (4 mmol/L – 7.8 mmol/L) before having a meal and should be less than 180 mg/dL within 1 – 2 hours after a meal [5]. Diabetics might experience hyperglycemia (> 230 mg/dL) or hypoglycemia (< 65 mg/dL) if the glucose level is not monitored and controlled accordingly. Invasive and minimally-invasive methods have been widely used in the market to check the blood glucose levels despite the many limitations and discomforts, particularly, the economic cost of the disposable lancet and strips used to draw and test the blood samples [6].

Researchers have conducted thorough investigations over the last decade on alternative, non-invasive methods of blood glucose monitoring. Among those methods, the microwave sensing has shown to be promising due the penetration ability of the EM waves. Such techniques rely on the fact that the glucose level in a person's blood affects the EM properties of the blood (dielectric constant and loss tangent) [7] and thus seek for a promising correlation between the measured EM properties of blood and its glucose content. A popular method for measuring the complex permittivity for liquids is through using an open-ended coaxial probe [8]. Such probes can be used for measurements in a convenient laboratory setup but not good candidates for non-invasive sensing due their high cost, massive bulky nature and confined penetration depth [9-10]. On the other hand, we find the microwave resonator sensors more attractive and offer a great advantage for measuring the complex permittivity for tested solids and liquids more suitably using low power with a flexibility in design and fabrication process that is both simple and cost-effective. Additionally, these resonators feature the non-invasive measurements in nondestructive fashion. Microwave resonators have specific resonance characteristics (resonance frequency, bandwidth and quality coefficient) based on their physical dimensions and the dielectric properties of the substrate. The sensing method is based on the modification of the resonance peaks and/or frequency when in the presence of the sample to be characterized, called the superstrate. This superstrate (e.g. glucose sample) is placed on the surface of the resonator in a region where the electric-field is highly concentrated to achieve the most interaction between the radiated field and the tested tissue at the operating frequency. The dielectric properties of tissues are calculated indirectly by analyzing these changes in the resonance behavior [11-13].

In this paper, we propose a microwave bio-sensor that consists of a triple-poles complementary split ring resonator (CSRR) coupled to a planar microstrip line etched on top of a dielectric substrate. The designed sensor is simulated for sensing the glucose levels non-invasively in aqueous solutions of concentrations in the range (70–120 mg/dL) that are characterized by the high permittivity and dielectric losses in the frequency range (1 – 10 GHz).

## II. Dielectric Properties of Glucose Solutions

For preliminary investigations on non-invasive glucose detection using RF sensors, aqueous glucose solutions can be utilized in the RF experiments to mimic the blood behavior at disparate glucose concentrations and assure sufficient repeatability of the measurements. This approximation is valid since water contributes high percentage (≈ 50% of total volume) of the whole human-beings blood [14]. Aqueous solutions introduce dispersion to the EM waves propagated through their media at different frequencies. The dielectric response in the frequency domain of tissues having high water content can be characterized by the Debye Relaxation Model [15], which

This work was supported by the CREATE NSERC grant, Erasmus+ exchange grant, and by Schlegel UW Research Institute of Aging (UW-RIA).

describes the reorientation of molecules that could involve translational and rotational diffusion, hydrogen bond, and structural arrangement. The relative permittivity for aqueous solutions is designated as a function of the angular frequency $\omega = 2\pi f$ and the concentration $\xi$ of any hydrophilic substance by

$$\epsilon_r(\omega, \xi) = \epsilon_r'(\omega, \xi) - j\epsilon_r''(\omega, \xi) \quad (1)$$

where $\epsilon_r'$ is the dielectric constant that represents the energy stored in the material when exposed to an external electric-field and $\epsilon_r''$ symbolizes the dielectric loss as given in (2) for the total absorbed energy due the ionic conduction $\epsilon_{r\sigma}''$ and dipole rotation $\epsilon_{rd}''$ loss mechanisms. The ratio between dissipated and stored energy is the loss tangent $tan\delta$ [16].

$$\epsilon_r''(\omega, \xi) = \epsilon_{rd}'' + \epsilon_{r\sigma}'' = \epsilon_{rd}''(\omega, \xi) + \frac{\sigma}{\omega\epsilon_o} \quad (2)$$

where $\sigma$ is the ionic conductivity of the material in S/m and $\epsilon_o$ is the reference free space permittivity (8.854x10$^{-12}$ F/m) [14]. According to the first order Debye relaxation model, the complex permittivity $\epsilon_r$ of a dispersive material, such as glucose-water solutions, can be calculated using Eq. (3) with the knowledge of four parameters denoted as static permittivity $\epsilon_{stat}$, infinity permittivity $\epsilon_\infty$, relaxation time $\tau$, and the static conductivity $\sigma_s$ [17]. However, the last term can be neglected for low-conductive materials

$$\epsilon_r(w, \xi) = \epsilon_\infty(\xi) + \left(\frac{\epsilon_{stat}(\xi) - \epsilon_\infty(\xi)}{1 + j\omega\ (\xi)}\right) + \frac{\sigma_s}{j\omega\epsilon_o} \quad (3)$$

Many studies have focused on extracting permittivity measurements for liquid samples of relatively high glucose concentrations at low frequencies using different setups; antennas [18-19], resonant cavities [20-21], open waveguides [22], split-ring dielectric resonators [7][23], open-ended coaxial probes [24-27]. In this study we adopted the Debye model parameters proposed by Hofmann et al in [26-27] where the influence of increasing glucose concentrations in aqueous solutions is investigated by measuring the complex permittivity of glucose-water solutions of 50, 250, 1000, and 2000 mg/dL using a commercial coaxial probe kit connected to VNA in a frequency range up to 40 GHz. The permittivity dependency in concentrations was modelled by fitting the collected data using a single-pole Debye model with the following parameters

$$\epsilon_\infty(\xi) \approx 5.38 + (30 \times 10^{-3}).\xi \quad (4)$$
$$\epsilon_s(\xi) \approx 80.68 - (0.207 \times 10^{-3}).\xi \quad (5)$$
$$\tau(\xi) \approx 9.68 + (0.23 \times 10^{-3}).\xi \quad (6)$$

The model parameters for our concentrations of interest are computed in Table I. Figures 1(a) and 1(b) plot the dielectric constant and loss tangent for the various concentrations over the frequency range 1-10 GHz. We can notice a slight increase in the dielectric constant and decrease in loss tangent with increasing the glucose concentrations like the pattern in [9].

Table I: Extracted Debye coefficients for each glucose concentration

| Glucose samples (mg/dL) | $\epsilon_\infty$ | $\epsilon_s$ | $\tau$ (ps) |
|---|---|---|---|
| 70 | 7.48 | 80.666 | 9.696 |
| 80 | 7.78 | 80.663 | 9.698 |
| 90 | 8.08 | 80.661 | 9.70 |
| 100 | 8.38 | 80.659 | 9.703 |
| 110 | 8.68 | 80.657 | 9.705 |
| 120 | 8.98 | 80.655 | 9.708 |

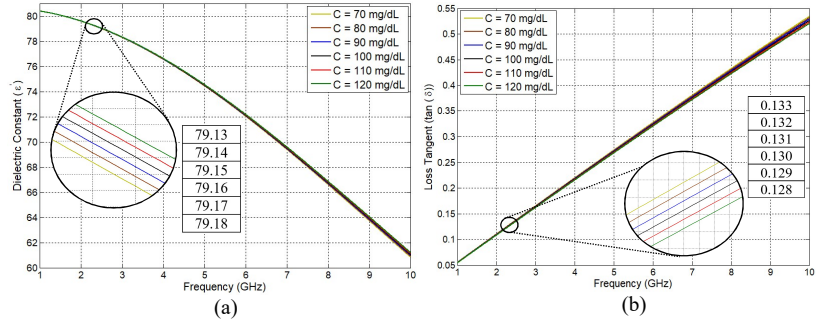

Fig. 1. Dielectric measurements of glucose solutions of different concentrations.

### III. THE PROPOSED BIO-SENSOR

The developed sensor is a circular ring resonator with localized elements. CSRR is one type of microwave resonators with smaller dimensions than the guided wavelength ($< \lambda_g/10$) and is featured by its high sensitivity to characterize the EM properties of materials [28-30]. It has a frequency-selective nature that resonates at a specific resonance frequency where the electric and magnetic energy stored in the frequency-dependent parasitic inductor and capacitor are equal. The resonance frequency $f_{rn} = nc/2\pi r\sqrt{\epsilon_e}$ depends on the geometrical parameters of the CSSR and the substrate specifications. CSRR consists of one or more slits in the form of concentric interrupted rings nested within each other and engraved in the plane of copper mass. Unlike the conventional SRR, the rings of the CSRR are dielectric and the slots are metallic. The CSRR behaves like an *RLC* resonant circuit; the winding of the rings acts as inductance *L*, the gaps and spacings of the rings create a parallel capacitance *C,* and the conductive and dielectric losses are modelled by a resistance *R*.

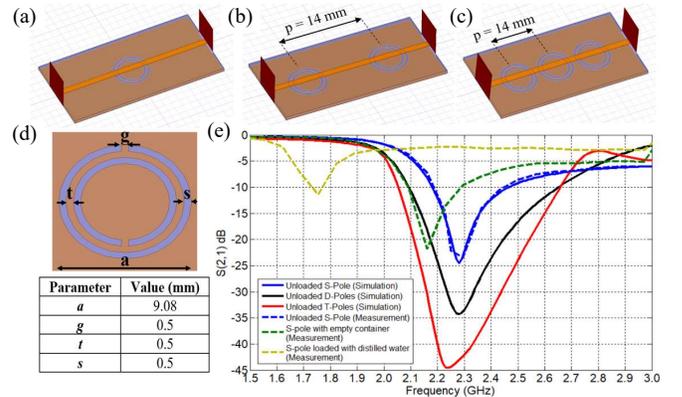

Fig. 2. CSRR (a) single-pole (b) double-poles (c) triple-poles (d) cell configuration and parameters (e) resonance characteristics.

The proposed CSRR is designed in HFSS in multiple-poles configurations as shown in Fig. 2(a), 2(b), and 2(c). The CSRR cell consists of two concentric split-rings engraved in the ground plane. The double split-ring are designed in the form of two circular-shaped loops with a gap between each as shown in Fig. 2(d). The resonator is coupled with the RF power supply via a copper microstrip line of 50 Ω and 40 mm length etched on the upper face of the FR4 substrate ($\epsilon_r' = 4.75, tan\delta = 0.02$, *h = 0.7 mm, W = 20 mm, L = 40 mm*). The access microstrip line was sized from ADS Line-Calc line synthesis module to have a characteristic impedance of 50 Ω at 2.29 GHz. A width of $W_{line}$ = 1.5 mm ribbon was obtained. The dimensions of the resonator have been optimized with the HFSS software to obtain a resonance frequency $f_0$ of 2.29 GHz as shown in Fig. 2(e) for the unloaded single-pole. Double and triple-poles achieve steeper Q-factors at 2.28 and 2.24 GHz, respectively, as shown. A shift of about 130 MHz is measured when the single-pole is loaded with an empty glass container, further shift of 405 MHz is recorded when a distilled water of 1.5 mL is loaded. The sensor is loaded with multiple layers as shown in Fig. 3, a slim glass layer of thickness $h_{glass} = 0.13$ mm is added on the surface of the resonator to avoid short-circuiting the slots of the CSRR when loaded by lossy glucose samples. This will introduce a few MHz shifts in resonance. In addition, to mimic the application of the bio-sensor for portable glucose monitoring when applied on top of a diabetic's hand; we add a skin layer of thickness $h_{skin} = 0.1$ mm ($\epsilon'_r = 38.1, tan\delta = 0.28$) [29]. Lastly, the glucose samples of different concentrations 70, 80, 90, 100, 110, and 120 mg/dL are loaded onto the sensing region in a rectangular shape of thickness $h_{glucose} = 2$ mm resulting in a frequency shift of about 1.08 GHz as shown in Fig. 4 that compares the $S_{21}$ response for the triple-poles CSRR against the conventional single and double-poles. It is noted that the TP-CSRR has steeper resonance depths that realize a better sensitivity for detecting the tiny variations in the loss tangent for different glucose concentrations. For instance, $S_{21}$ changes by 0.092 dB and 0.054 dB at $f_{r2}$ and $f_{r3}$, respectively, as the glucose concentration increases from 70 to 80 mg/dL. The sensitivity $\Delta S_{21}/\Delta C$ in dB/(mg/dL) is computed in Table III at the resonance frequencies for the multiple-poles configurations. In the case of the TP-CSRR, an intense electric-field is coupled to the sensing region where the lossy glucose sample acquires more interaction with the near-field, and hence induces notable variation in the intrinsic characteristics of the resonator for small changes in the EM properties of the glucose. Figure 5 depicts the $S_{21}$ responses (insertion losses) for the TP-CSRR when is loaded with a rectangular glucose sample of 70 mg/dL at different volumes (thicknesses). Apparently, the shift in resonance frequency is highly dependent on the volume of the glucose layer. With an increased volume we can notice a gradual decrease in the resonance peaks and a pattern of slight increasing in the resonance frequency shift that saturates at the largest volume of $V = 5.4$ mL. Furthermore, the results have shown that CSRR achieve higher sensitivity for loss in the glucose solutions when loaded with samples of smaller volumes as demonstrated for the case of $V = 0.54$ mL at which the resonator shows a sensitivity of about 0.055 dB/(mg/dL) at $f_{r3} = 4.94$ GHz as shown in the zoom-in window.

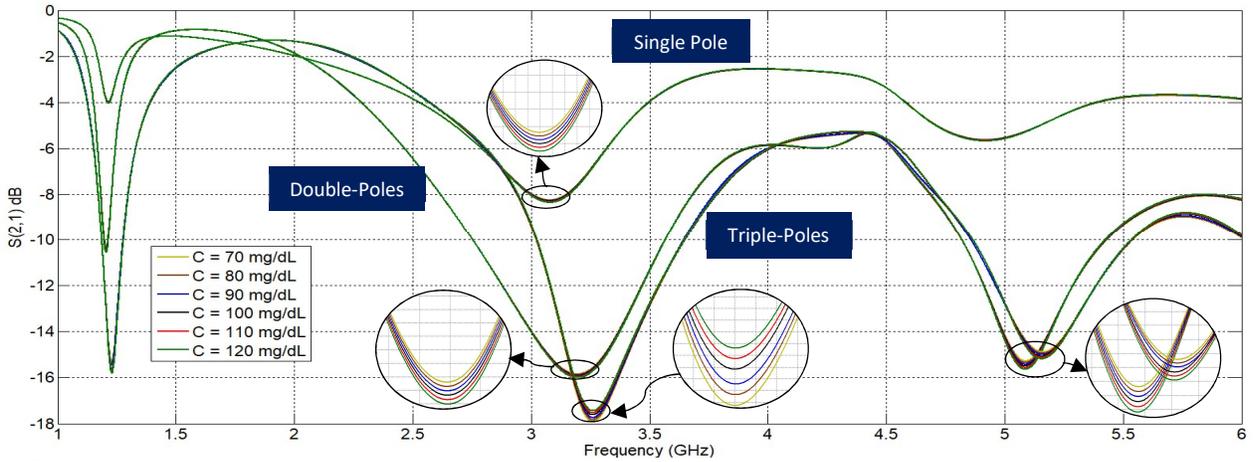

Fig. 4. Simulation results of the transmission coefficient $S_{21}$ for different glucose concentrations loaded on multiple-poles CSRR over the frequency range (1 – 6 GHz) with zoom-in windows for the resonance depth at harmonic resonances.

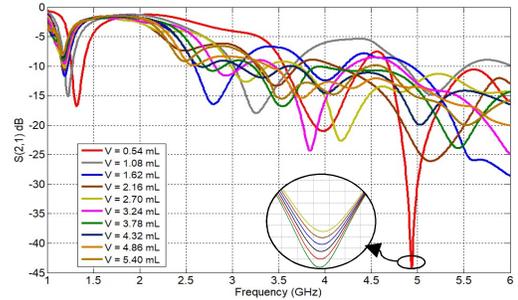

Fig. 5. Volume effect on the resonance frequency and depth for TP- CSRR when loaded with 70 mg/dL

Table III: Sensitivity comparisons for multiple-poles CSSR configurations.

| Single pole | | Double poles | | Triple poles | |
|---|---|---|---|---|---|
| $f_{rn}$ (GHz) | Sensitivity dB/(mg/dL) | $f_{rn}$ (GHz) | Sensitivity dB/(mg/dL) | $f_{rn}$ (GHz) | Sensitivity dB/(mg/dL) |
| 1.215 | 1.43x10$^{-3}$ | 1.22 | 4.77x10$^{-3}$ | 1.233 | 6x10$^{-3}$ |
| 3.08 | 2.18x10$^{-3}$ | 3.195 | 3.25x10$^{-3}$ | 3.262 | 9.16x10$^{-3}$ |
| 4.92 | 1.11x10$^{-3}$ | 5.088 | 6.98x10$^{-3}$ | 5.165 | 5.57x10$^{-3}$ |

## CONCLUSION

In this paper, we propose a non-invasive microwave glucose sensor that consists of triple-poles complementary split ring resonators coupled to a planar microstrip line on top of a dielectric substrate. The sensor is simulated for detecting the glucose in aqueous solutions of concentrations in the diabetics' range (70–120 mg/dL). Results have shown that the second and third harmonic resonances exhibit higher sensitivity for glucose detection up to 9.16x10$^{-3}$ dB/(mg/dL). This sensitivity is highly dependent on the sample volume with a better distinguishability for smaller volumes. Consequently, such CSRR can be used to sense the volume of a loaded dielectric layer from the shift in resonant frequency. The proposed CSRR is under fabrication and the results will be verified via VNA measurements next.

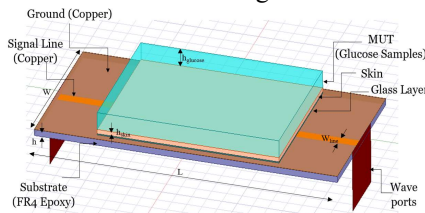

Fig. 3. The sensor structure loaded with glucose samples